# A Parallel Simulator for Massive Reservoir Models Utilizing Distributed-Memory Parallel Systems

Hui Liu [1],[*] Lihua Shen[1], Yan Chen[1,2][*], Kun Wang[1], Bo Yang[1], Zhangxin Chen[1][*]

[1] Department of Chemical and Petroleum Engineering
University of Calgary, 2500 University Drive NW, Calgary, AB, Canada
[2] College of Mathematics and informatics
South China Agricultural University, Guangzhou 510642, China

**Abstract**

This paper presents our work on developing parallel computational methods for two-phase flow on modern parallel computers, where techniques for linear solvers and nonlinear methods are studied and the standard and inexact Newton methods are investigated. A multi-stage preconditioner for two-phase flow is applied and advanced matrix processing strategies are studied. A local reordering method is developed to speed the solution of linear systems. Numerical experiments show that these computational methods are effective and scalable, and are capable of computing large-scale reservoir simulation problems using thousands of CPU cores on parallel computers. The nonlinear techniques, preconditioner and matrix processing strategies can also be applied to three-phase black oil, compositional and thermal models.

**Keywords:** Multi-phase flow, oil-water, porous media, simulation, parallel computer

# INTRODUCTION

Large-scale geological models in petroleum reservoirs and their field-scale simulation models with millions of grid blocks (cells, elements) are being applied to represent complex geological heterogeneity and to capture their high-resolution phenomena, which lead to extremely long simulation run time because of difficulties from solutions of linear and nonlinear systems with a huge number of grid blocks. Nowadays, most of available commercial reservoir simulators have been developed for personal computers and workstations, which are limited by their memory size, memory speed and CPU performance. When simulating large-scale reservoir models, it may take days or even longer to complete a simulation run. Parallel reservoir simulators and efficient numerical methods for parallel computers are essential to improve simulation runs and simulators' capability.

---

[*]Authors to whom correspondence may be addressed. Email addresses: liu1@ucalgary.ca, cheny@scau.edu.cn, zhachen@ucalgary.ca

Reservoir simulations have been studied for decades by researchers, and various techniques have been developed to model new recovery processes and to accelerate computer simulations, including new reservoir models and numerical methods[27,28], nonlinear methods[23,28], linear solvers and preconditioner techniques[17,11,12,13,14,15], and parallel computing[1,2]. Shiralkar and his collaborators[4] developed a parallel simulator, FALCON, using FORTRAN 90 and high performance FORTRAN (HPF), which could run on a variety of computing platforms, such as Cray T3D and T3E, SGI Origin 2000, Thinking Machines CM5 and IBM RS6000. Killough and Bhogeswara applied local refinement methods to reservoir simulations and designed effective preconditioners for parallel linear solvers in their parallel simulator[7]. The parallel simulator developed by Parashar et al. successfully handled multiple fault blocks with multiple physics[5]. Rutledge and his collaborators implemented a parallel compositional simulator for SIMD computers, which solved pressure unknowns implicitly and saturation unknowns explicitly[3]. Kaarstad et al. presented a two-dimensional two-phase (oil and water) parallel reservoir simulator, which could solve large problems with up to one million grid blocks[6]. Dogru et al. developed the parallel black oil simulator, which could calculate models with a billion grid blocks[8,9]. Zhang et al. developed a parallel platform for adaptive finite element methods and finite volume methods[35,25], which was also used to develop a parallel black oil simulator using discontinuous Galerkin methods. Linear systems from reservoir simulations are usually ill-conditioned, especially when heterogeneous geology is applied. The Krylov subspace linear solvers and multi-grid solvers have been widely adopted in reservoir simulations. Advanced preconditioner techniques have also been proposed and applied to reservoir simulations, including point-wise and block-wise incomplete factorization[11,12], domain decomposition[26], constrained pressure residual (CPR)[13,14,22], multi-stage[20] and FASP methods[21].

This paper deals with developing fast numerical simulations for two-phase flow on distributed memory parallel computers. A finite difference method is employed to compute a two-phase model due to its simplicity. Due to the high nonlinearity of this model, the standard Newton (or Newton-Raphson) method and the inexact Newton method[23] are both investigated. The standard Newton method tends to solve linear systems as accurately as possible, in which the solutions may be expensive to obtain when their condition numbers are large. In contrast, the inexact Newton method just solves linear systems approximately and the stop criterion for each linear system changes depending on its convergence history. For the linear systems derived from the (inexact) Newton methods, they are usually ill-conditioned, especially when the geological models are highly heterogeneous and the sizes of grid blocks of these models are large, which are difficult to solve. An efficient multi-stage preconditioner[22] is applied to these linear systems, which is based on the classical constrained pressure residual method (CPR). It solves the pressure equation using the AMG (algebraic multigrid) method to minimize errors of the solutions to linear systems and the entire linear system using the restricted additive Schwarz (RAS) method. The AMG method and the RAS method have excellent scalability so this preconditioner has excellent scalability. In addition, advanced matrix processing techniques are also studied. Two decoupling strategies are applied to linear systems, which are the Quasi-IMPES (implicit pressure and explicit saturation) strategy and the alternative block factorization (ABF) strategy. A local potential reordering technique is also developed, which reorders the unknowns by the potential of grid blocks. Although the techniques are applied for a two-phase flow model, they are also efficient for three-phase black oil and compositional models. The nonlinear methods, the domain decomposition preconditioner, and matrix processing strategies can even be applied to a thermal model. Numerical experiments show that these techniques are efficient and scalable for parallel computing.

# TWO-PHASE OIL-WATER FLOW MODEL

The two-phase model assumes that the simultaneous flow has two phases (oil and water) and two components (oil and water), and they are immiscible. It can also describe a case when solution gas stays in the oil phase during the entire simulation period. The model also assumes temperature never changes.

Darcy's law describes the flow of fluids through a porous media, which establishes a relationship between the volumetric flow rate and the pressure gradient,

$$Q = -\frac{KA\Delta p}{\mu L}, \tag{1}$$

where $K$ is the permeability of a given reservoir, $A$ is the area in the flow direction, $\Delta p$ is the pressure difference, $\mu$ is the viscosity of the fluid, and $L$ is the length of the reservoir. In three-dimensional space, its differential form law is

$$q = \frac{Q}{A} = -\frac{K}{\mu}\nabla p. \tag{2}$$

By combining Darcy's law, mass conservation law and gravitational force, the two-phase model[27] is described as the following equation,

$$\begin{cases} \frac{\partial}{\partial t}(\phi s_o \rho_o) = \nabla \cdot (\frac{KK_{ro}}{\mu_o}\rho_o \nabla \Phi_o) + q_o \\ \frac{\partial}{\partial t}(\phi s_w \rho_w) = \nabla \cdot (\frac{KK_{rw}}{\mu_w}\rho_w \nabla \Phi_w) + q_w. \end{cases} \tag{3}$$

A no-flow condition is adopted as the boundary condition.

$\phi$ is the porosity of the reservoir, which is computed as,

$$\phi = \phi_r(1 + c_r(p_o - p_r)), \tag{4}$$

where $c_r$ is the rock compressibility factor and $p_r$ is the reference pressure of rock.

Let $\alpha$ ($\alpha = o, w$) be a phase. $\Phi_\alpha$ is phase potential (total absolute pressure), which is defined as,

$$\Phi_\alpha = p_\alpha + \rho_\alpha g z, \tag{5}$$

where $p_\alpha$ is phase pressure, $\rho_\alpha$ is phase density, $g$ is the standard gravity, and $z$ is the reservoir depth. The density is a function of pressure, which is written as,

$$\rho_\alpha = \rho_\alpha^r(1 + c_{\alpha,r}(p_\alpha - p_{\alpha,r})). \tag{6}$$

Here, $\rho_\alpha^r$ is the reference density at reference pressure $p_{\alpha,r}$, and $c_{\alpha,r}$ is the reference compressibility factor. $\mu_\alpha$ is the viscosity of a phase, which is calculated as

$$\mu_\alpha = \mu_{\alpha,r} + c_{\mu,\alpha}(p_\alpha - p_{\alpha,r}), \tag{7}$$

where $\mu_{\alpha,r}$ is the reference viscosity at reference pressure $p_{\alpha,r}$ and $c_{\mu,\alpha}$ is the factor for pressure dependence of the phase viscosity.

In this model, oil and water occupy the reservoir. Their volumes are described by oil saturation, $s_o$, and water saturation, $s_w$, which satisfy,

$$s_o + s_w = 1. \tag{8}$$

The oil phase pressure, $p_w$, and the water phase pressure, $p_o$, may be different due to capillary pressure $p_c$. They are related as,

$$p_w = p_o - p_c(s_w). \tag{9}$$

The capillary pressure $p_c$ is a function of water saturation and is user input. $K_{r\alpha}$ is the relative permeability. $K_{rw}$ and $K_{ro}$ are functions of water saturation, $s_w$. $K$ is the permeability, which has three components in $x$, $y$ and $z$ directions, noted as $K_x$, $K_y$ and $K_z$.

The Peaceman method is adopted for well modeling. A well may have many perforations. For each perforation at a grid block $m$, its well rate $q_{\alpha,m}$ is calculated by the following formula,

$$q_{\alpha,m} = W_i \frac{\rho_\alpha K_{r\alpha}}{\mu_\alpha}(p_b - p_\alpha - \rho_\alpha g(z_b - z)), \tag{10}$$

where $p_b$ is the bottom hole pressure defined at the reference depth $z_b$, $W_i$ is the well index, $z$ is the depth of the perforation in grid block $m$, and $p_\alpha$ is phase pressure in grid block $m$. $W_i$, which is the well index, defines the relationship among the well bottom hole pressure, flow rate, and grid block pressure[27].

Various operation constraints may be applied to a well at different time stage, such as a fixed bottom hole pressure, a fixed oil rate, a fixed water rate or a fixed liquid rate. When the fixed bottom hole pressure condition is applied to some well, the constraint equation is,

$$p_b = c, \tag{11}$$

where $c$ is a constant and known. The fixed water rate condition is the following equation:

$$\sum_m q_{w,m} = q_w, \tag{12}$$

where $q_w$ is constant and known. For the fixed oil rate condition, its constraint equation is

$$\sum_m q_{o,m} = q_o, \tag{13}$$

where $q_o$ is fixed. For the fixed liquid rate condition, the constraint equation is

$$\sum_m (q_{o,m} + q_{w,m}) = q_o + q_w. \tag{14}$$

## NUMERICAL METHODS

To solve equation (3), the fully implicit method (FIM) is applied, and the oil phase pressure $p$, water saturation $s_w$ and the well bottom hole pressure $p_b$ are chosen as the unknowns. The time differentiation term is discretized by the backward Euler difference scheme and the space differentiation terms are discretized by the cell-centered finite difference method[27].

Let $u^n$ be the value of the function $u$ at time step $n$, then its derivative at time step $(n+1)$ can be approximated by backward difference,

$$\left(\frac{\partial u}{\partial t}\right)^{n+1} = \frac{u^{n+1} - u^n}{\Delta t}, \tag{15}$$

where $\Delta t$ is time step. If we assume that $d$ ($d = x, y, z$) is any space direction and $A$ is the area of the faces along $d$ direction in a grid block, then the transmissibility term $T_{\alpha,d}$ of phase $\alpha$ is defined as

$$T_{\alpha,d} = \frac{KK_{r\alpha}}{\mu_\alpha} \rho_\alpha \frac{A}{\Delta d}, \tag{16}$$

where $\Delta d$ is the cell length along $d$ direction. Then the nonlinear system is written as,

$$\begin{cases} \frac{V}{\Delta t}[(\phi \rho_o s_o)^{n+1} - (\phi \rho_o s_o)^n] &= \nabla \cdot (T_o^{n+1} \nabla \Phi_o^{n+1}) + q_o^{n+1} \\ \frac{V}{\Delta t}[(\phi \rho_w s_w)^{n+1} - (\phi \rho_w s_w)^n] &= \nabla \cdot (T_w^{n+1} \nabla \Phi_w^{n+1}) + q_w^{n+1} \end{cases} \tag{17}$$

where $V$ is the volume of a specific grid block.

The transmissibility term is defined on each face of a cell. Therefore, a cell has six items defined. For any face shared by two cells, the value of the transmissibility term is the same in each cell. In this case, the mass of any component is conservative. In this paper, the upstream technique is employed to compute the transmissibility term. For a grid block $(i, j, k)$, the transmissibility term in $x$ direction, $(T_{\alpha,x})_{i \pm \frac{1}{2}, j, k}$, is defined as

$$(T_{\alpha,x})_{i \pm \frac{1}{2}, j, k} = \begin{cases} (T_{\alpha,x})_{i \pm 1, j, k}, & \text{if } \Phi_{i \pm 1, j, k} \geq \Phi_{i, j, k} \\ (T_{\alpha,x})_{i, j, k}, & \text{if } \Phi_{i \pm 1, j, k} < \Phi_{i, j, k} \end{cases}, \tag{18}$$

whose value is from the cell with larger phase potential. Transmissibility terms in $y$ and $z$ directions are defined the same way.

We assume a grid has $n$ blocks (cells, elements) and $\tau$ wells exist in the model. The final nonlinear system is represented by

$$F(\vec{x}) = 0, \tag{19}$$

where

$$\vec{x} = \begin{pmatrix} \vec{p} \\ \vec{s_w} \\ \vec{p_b} \end{pmatrix}, \tag{20}$$

$$\vec{p} = \begin{pmatrix} p_1 \\ p_2 \\ \ldots \\ p_n \end{pmatrix}, \tag{21}$$

$$\vec{s_w} = \begin{pmatrix} s_{w1} \\ s_{w2} \\ \ldots \\ s_{wn} \end{pmatrix}, \tag{22}$$

and

$$\vec{p_b} = \begin{pmatrix} p_{b1} \\ p_{b2} \\ \ldots \\ p_{b\tau} \end{pmatrix}. \tag{23}$$

$F$ is a nonlinear mapping from $\mathbb{R}^N$ to $\mathbb{R}^N$ ($N = 2 \times n + \tau$). The nonlinear system is solved by the Newton method (or inexact Newton method). The linear systems are also an order of $N$. Here we mention that the properties related to saturation are strongly nonlinear while the properties related to pressure are weakly nonlinear.

## Nonlinear Methods

In the reservoir simulations, a set of nonlinear equations are required to be solved at each time step, and the objective is to find an $x^* \in \mathbb{R}^N$ such that $\|F(x^*)\| < \varepsilon$, where $\varepsilon$ is the stopping criterion (tolerance) for nonlinear methods. Traditionally, the standard Newton method is applied to the nonlinear system (19), and the stopping tolerance, $\varepsilon$, is small, such as $10^{-5}$.

It is well-known that linear systems from reservoir simulations are hard to solve, and the solution time may occupy more than 70% of the total simulation time[27]. To obtain enough accuracy, the stopping tolerance for the standard Newton methods is set to be small, which can make the situation even worse. In practice, we know that in the first few Newton iterations, the solution $x^l$ may be far from the real solution $x^*$. Therefore, it may not be necessary to solve the linear systems accurately[23].

The inexact Newton method is an extension of the standard Newton method, which satisfies the following relationship during the solution procedure:

$$\|Ay - b\| \leq \eta_l \|b\|, \tag{24}$$

where $A$ is the Jacobian matrix and $y$ is only an approximate solution depending on the parameter $\eta_l$. The algorithm for the inexact Newton method is described in Algorithm 1[23].

---
**Algorithm 1** The Inexact Newton Method
---
1: Given an initial guess $x^0$ and stopping criterion $\varepsilon$, let $l = 0$ and assemble the right-hand side $b$.
2: **while** $\|b\| \geq \varepsilon$ **do**
3:     Assemble the Jacobian matrix $A$.
4:     Determine $\eta_l$.
5:     Find $y$ such that
$$\|Ay - b\| \leq \eta_l \|b\|, \tag{25}$$
6:     Let $l = l + 1$ and $x^l = x^{l-1} + y$.
7: **end while**
8: $x^* = x^l$ is the approximate solution of the nonlinear system, $F(x) = 0$.

---

$\eta_l$ is a forcing term, which forces the residual of the linear system $Ay = b$ to be small. The choice of the forcing term is designed to achieve fast local convergence and to avoid over-solving the linear system[23], three of which are listed as follows:

$$\eta_l = \begin{cases} \dfrac{\|b(x^l) - r^{l-1}\|}{\|b(l-1)\|}, \\[6pt] \dfrac{\|b(x^l)\| - \|r^{l-1}\|}{\|b(x^{l-1})\|}, \\[6pt] \gamma\left(\dfrac{\|b(x^l)\|}{\|b(x^{l-1})\|}\right)^\beta, \end{cases} \quad (26)$$

where $r^l$ is the residual of the $l$-th iteration,

$$r^l = b - Ay. \quad (27)$$

A typical value for $\beta$ is $\frac{\sqrt{5}+1}{2}$. The second one is employed in this paper. Besides, $\eta_k$ is forced to satisfy: $\eta_k \in [0.01, 0.1]$.

## Solution of Linear Systems

Since the pressure unknown $\vec{p}$ is placed before the saturation unknown $\vec{s_w}$ and bottom hole pressure $\vec{p_b}$, the matrix $A$ has the following structure:

$$A = \begin{pmatrix} A_{pp} & A_{ps} & A_{p\sigma} \\ A_{sp} & A_{ss} & A_{s\sigma} \\ A_{\sigma p} & A_{\sigma s} & A_{\sigma\sigma} \end{pmatrix}, \quad (28)$$

where the sub-matrices $A_{pp}$, $A_{ss}$, and $A_{\sigma\sigma}$ are the matrices corresponding to the pressure unknowns, the saturation unknowns, and the bottom hole pressure unknowns respectively. Other sub-matrices are coupled items.

In this paper, the matrix $A$ is decoupled first to weaken the relationship between the pressure and saturation unknowns, which is described as,

$$D^{-1}Ax = D^{-1}b, \quad (29)$$

where $D$ is the decoupling operator (matrix). Then a potential reordering technique is applied. In the end, a multi-stage preconditioner are employed to work with Krylov solvers.

## Matrix Decoupling Strategy

In order to weaken the strong coupling between the pressure unknowns and the saturation unknowns, a decoupling operator is necessary to the Jacobian matrix $A$. The decoupling operator should be computationally cheap and effective. For the two-phase model, the Quasi-IMPES strategy[18] and the alternative block factorization (ABF) strategy[15] are studied. A detailed study on these decoupling strategies can be found in[20].

The ABF method is written as

$$D_{ABF} = \begin{pmatrix} \mathbf{Diag}(A_{pp}) & \mathbf{Diag}(A_{ps}) & 0 \\ \mathbf{Diag}(A_{sp}) & \mathbf{Diag}(A_{ss}) & 0 \\ 0 & 0 & I_\sigma \end{pmatrix},$$

where **Diag**($B$) stands for the diagonal of matrix $B$. The ABF method applies to the reservoir blocks only and the well blocks keep unchanged. The Quasi-IMPES method is defined as follows:

$$D_{QI} = \begin{pmatrix} I_n & \textbf{Diag}(A_{ps})\textbf{Diag}(A_{ss})^{-1} & 0 \\ 0 & I_n & 0 \\ 0 & 0 & I_\sigma \end{pmatrix}.$$

The Quasi-IMPES decoupling strategy is simpler and less aggressive than the ABF strategy. It preserves the saturation and well blocks. $A_{sp}$ contributes to the pressure blocks.

The new linear system is equivalent to the original linear system $Ax = b$. Therefore, it is also denoted as $Ax = b$. For parallel computing, the matrix-matrix multiplication operation $D^{-1}A$ and the matrix-vector operation $D^{-1}b$ are local, and there is no communication involved. We can see that the scalability of the matrix decoupling strategies (the Quasi-IMPES strategy and the ABF strategy) is ideal.

## Local Potential Reordering Technique

For the development of linear solvers, it is well-known that proper matrix reordering techniques improve the efficiency of linear solvers and preconditioners. Some of the techniques reduce the bandwidth of a given matrix, such as the RCM (reverse Cuthill-MaKee) reordering method. Some of the techniques reduce fill-ins of the ILU methods, such as the MD (minimum degree) reordering method. Proper reordering techniques are important to the efficiency of linear solvers and preconditioners.

For reservoir simulation, upstream schemes are widely applied, which are determined by the potential of a phase, such as the oil phase. Kwok and Tchelepi developed a potential-based reordering strategy and a reduced Newton method was proposed based on this reordering technique[28]. In this paper, this strategy is applied to reorder the pressure unknowns and the saturation unknowns. The oil phase potential, $\Phi_o = p_o + \rho_o gz$, is employed for each grid block. For the $i$th grid block, it has a potential $\Phi_i$. Then $\Phi_1$, $\Phi_2$, $\cdots$, and $\Phi_n$ form a set $\Phi = \{\Phi_1, \Phi_2, \cdots, \Phi_n\}$.

Let us sort the set $\Phi$ in a descending order based on the values of $\Phi_i$ ($1 \leq i \leq n$). Then, for any $\Phi_i$, its location in the sorted set $\Phi$ is determined, say $L_i$, which is numbered from the left side. Now, we can define a mapping $Pm$ such that

$$Pm(i) = L_i. \qquad (30)$$

An example is given to demonstrate the definition of the mapping $Pm$. Assume that we have four grid blocks, their potentials are $\Phi_1 = 2.1$, $\Phi_2 = 1.3$, $\Phi_3 = 4.2$ and $\Phi_4 = 1$, and the sorted set $\Phi$ is $\{4.2, 2.1, 1.3, 1\}$. Then we have $Pm(1) = 2$ ($\Phi_1$), $Pm(2) = 3$ ($\Phi_2$), $Pm(3) = 1$ ($\Phi_3$), and $Pm(4) = 4$ ($\Phi_4$). We can see that the mapping $Pm$ defines a permutation of the set $\{1, 2, \cdots, n\}$.

We know that the unknown $x$ has the following form:

$$\vec{x} = \begin{pmatrix} \vec{p} \\ \vec{s_w} \\ \vec{p_b} \end{pmatrix}, \qquad (31)$$

whose dimension is $(2 \times n + \tau)$. Here, the pressure unknowns are numbered first, followed by the water saturation unknowns and the bottom hole pressure unknowns. Now, let us define the mapping for the whole linear system:

$$Pt(i) = \begin{cases} L_i, & 1 \leq i \leq n, \\ n + L_{i-n}, & n+1 \leq i \leq 2 \times n, \\ i, & 2 \times n + 1 \leq i \leq 2 \times n + \tau. \end{cases} \qquad (32)$$

Then the mapping $Pt$ defines a permutation of the set $\{1,2,3,\cdots,2\times n+\tau\}$. If we apply this mapping to the linear system $Ax = b$, the matrix $\tilde{A}$ of the reordered linear system still has the same structure as $A$:

$$\tilde{A} = \begin{pmatrix} \widetilde{A_{pp}} & \widetilde{A_{ps}} & \widetilde{A_{p\sigma}} \\ \widetilde{A_{sp}} & \widetilde{A_{ss}} & \widetilde{A_{s\sigma}} \\ \widetilde{A_{\sigma p}} & \widetilde{A_{\sigma s}} & \widetilde{A_{\sigma\sigma}} \end{pmatrix}. \tag{33}$$

The reason that we use the potential reordering technique is that the upstream schemes are applied and the fluid flows from grid blocks with high potential to grid blocks with low potential. After the reordering, the saturation block $\widetilde{A_{ss}}$ is close to a lower triangular matrix, which can be solved effectively by ILU methods. The ILU methods are more efficient for the reordered linear systems.

A global reordering for distributed parallel systems is extremely difficult, which involves heavy communications. Therefore, the reordering strategy is only applied to the local grid on each processor. In this case, there is no communication during the reordering process except information exchange required to define the mapping $Pt$. The communication volume is small and the strategy is scalable to parallel computing. The reordering technique here is called the local potential reordering technique.

## Multi-stage Preconditioner

The pressure equation of the two-phase model is elliptic (or parabolic) and the pressure unknowns contribute most of an error to this model. It is well-known that the algebraic multigrid methods (AMG) are the most effective methods to positive definite matrices. Wallis et al. introduced the classical constraint pressure residual (CPR) method to oil-water model and black oil model.

To introduce our multi-stage preconditioner, some notation is required. We have introduced the $x$, the pressure unknown $\vec{p}$, the saturation unknown $\vec{s_w}$ and the bottom hole pressure unknown $\vec{p_b}$. A global restriction operator for the pressure unknown is defined as

$$\Pi_r x = \vec{p}. \tag{34}$$

A prolongation operator $\Pi_p$ is defined as

$$\Pi_p \vec{p} = \begin{pmatrix} \vec{p} \\ \vec{0} \\ \vec{0} \end{pmatrix}. \tag{35}$$

$\Pi_p p$ has the same dimension as $x$.

We define the notation $\mathscr{M}(B)^{-1}b$ to represent the solution $y$ for linear system $By = b$ from an AMG solver. If the system $By = b$ is solved by the restricted additive Schwarz (RAS) method, we use the notation $\mathscr{R}(M)^{-1}b$ to represent the solution $y$. A CPR-like preconditioner, denoted by CPR-FPF preconditioner[22], has been designed, which is described in Algorithm 2. The method is a three-stage preconditioner. It is suitable for two-phase oil-water model, black oil model, compositional model and extended black oil models. We should mention that the RAS method and AMG method are scalable, so the CPR-FPF preconditioner is also scalable.

**Algorithm 2** The CPR-FPF Preconditioner
1: $x = \mathscr{R}(A)^{-1} f$.
2: $r = f - Ax$
3: $x = x + \Pi_p (\mathscr{M}(A_{pp})^{-1}(\Pi_r r))$.
4: $r = f - Ax$
5: $x = x + \mathscr{R}(A)^{-1} r$.

# NUMERICAL EXPERIMENTS

An IBM Blue Gene/Q system is applied to benchmark our reservoir models. It uses 64-bit PowerPC A2 processor, whose performance is low compared with Intel processors. However, it has a strong network relative to compute performance, which ensures its scalability. The IBM Blue Gene/Q is an excellent system for parallel computing. In the following sections, our application runs only one MPI processor on each CPU core. When we say using $n_p$ CPU cores, we mean that $n_p$ MPI processors are used.

## Validation

**Example 1** *This case is from CMG IMEX (mxsmo031.dat). The permeability varies from cell to cell, and the porosity varies in z direction. The mesh is $10 \times 10 \times 3$ with mesh size 1000ft. in x and y directions and 20, 30, 50 ft. in z direction from top to bottom. The depth of the top layer center is 926 ft. From top to bottom, the permeability in x direction is 500, 50, 200 mD, same as the permeability in y direction. The permeability in z direction is 60, 40, 20 mD. The porosity is 0.3.*

*Component properties: densities of gas, oil and water are 0.0647 lbm/ft3, 46.244 lbm/ft3 and 62.419 lbm/ft3.*

*The initial conditions are as follows: reference pressure 1000 psi at associated depth 100 ft, depth to water-oil contact is 2000 ft.*

*There are one injection well and one production well. Both are vertical. Injection well has maximum water injection rate 1000 bbl/day, maximum bottom hole pressure 2.0e+4 psi, well index 1.0e+5 with perforation at cell [1 1 1]. Production well has maximum oil production rate 2000 stb/day, maximum bottom hole pressure 100 psi. Well radius 0.25 ft with perforation at cell [10 10 3]. The well schedule is listed in Table 1.*

Table 1: The schedule of example two-phase model from mxsmo031.

| time(days) | injector(stw) | producer(sto) |
|:---:|:---:|:---:|
| 0 | 1000 | shutin |
| 1000 | 10000 | 10000 |
| 1200 | shutin | unchanged |
| 2000 | unchanged | 20000 |
| 8000 | 5000 | shutin |
| 9000 | shutin | 25000 |
| 17000 | 5000 | shutin |
| 20000 | stop | stop |

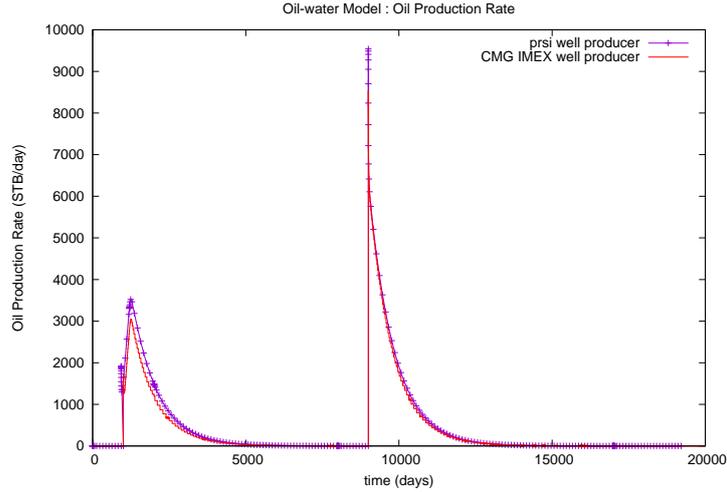

Figure 1: Oil production rate for well producer of example two-phase model from mxsmo031 (unit: stb/day).

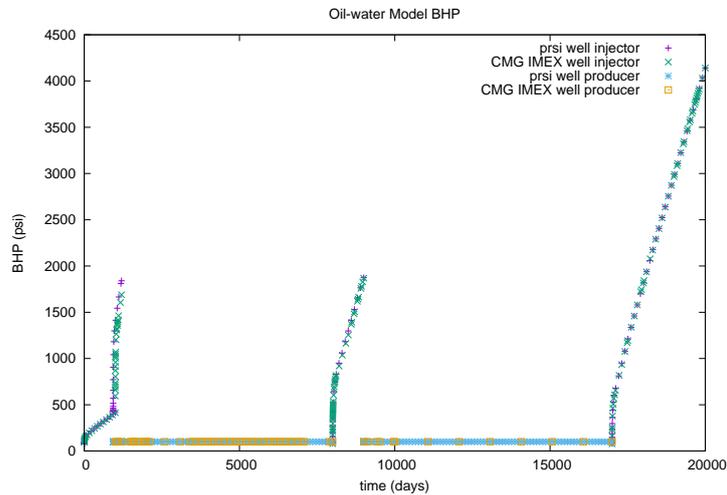

Figure 2: Oil production rate for well injector and producer of example two-phase model from mxsmo031 (pressure unit: psi).

Figure 1 shows oil production rate from our methods and CMG IMEX, from which we can see they match very well. Figure 2 compares bottom hole pressure from this paper and from CMG IMEX. Again, the results match very well. The same happens in Figure 3, which compares water rate with results from CMG IMEX. We can conclude that our methods and our implementation are correct.

**Example 2** *This case is also from CMG IMEX (mxspe1.dat). It is a 3D two-phase (oil-water) reservoir model with constant porosity and permeability varying in the z-direction layers. The mesh is 10× 10×3 with mesh size 1000 ft.in x and y directions and 20.0, 30.0, 50.0 ft. in z direction from top to bottom. The depth of the top layer center is 8335 ft. From top to bottom, the permeability in x direction is 500, 50, 200 mD, same as the permeability in y direction. The permeability in z direction is 60, 40, 20 mD. The porosity is 0.3.*

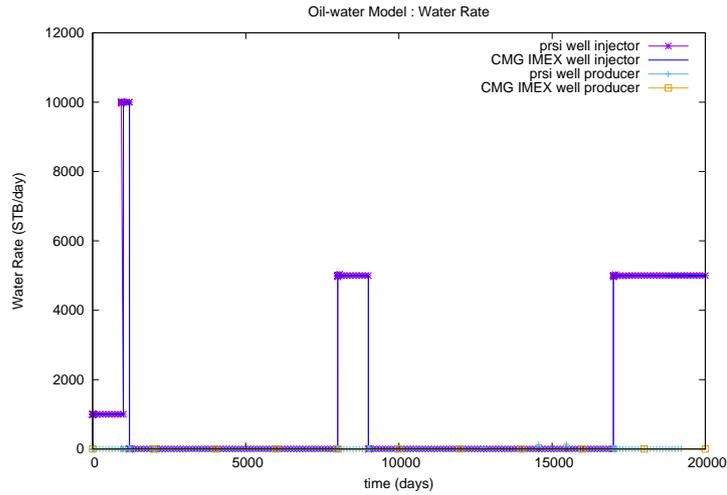

Figure 3: Water rate for well injector and producer of example two phase model from mxsmo031 (pressure unit: stb/day).

*Component properties: densities of gas, oil and water are 0.0647 lbm/ft3, 46.244 lbm/ft3 and 62.238 lbm/ft3.*

*The initial conditions are as follows: reference pressure 4800 psi at associated depth 8400 ft, depth to water-oil contact is 9500 ft.*

*There are one injection well and one production well. Both are vertical. Injection well has maximum water injection rate 1.0e+5 bbl/day, maximum bottom hole pressure 2.0e+4 psi, well index 1.0e+5 with perforation at cell [1 1 1]. Production well has maximum oil production rate 2.0e+4 stb/day, maximum bottom hole pressure 1000 psi. Well radius 0.25 ft with perforation at cell [10 10 3]. The simulation time is 3650 days.*

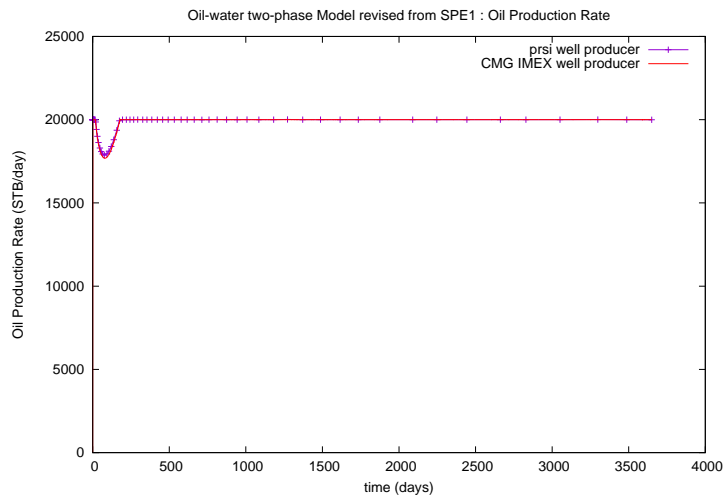

Figure 4: Oil production rate for well producer of example two-phase model from spe1 (unit: stb/day).

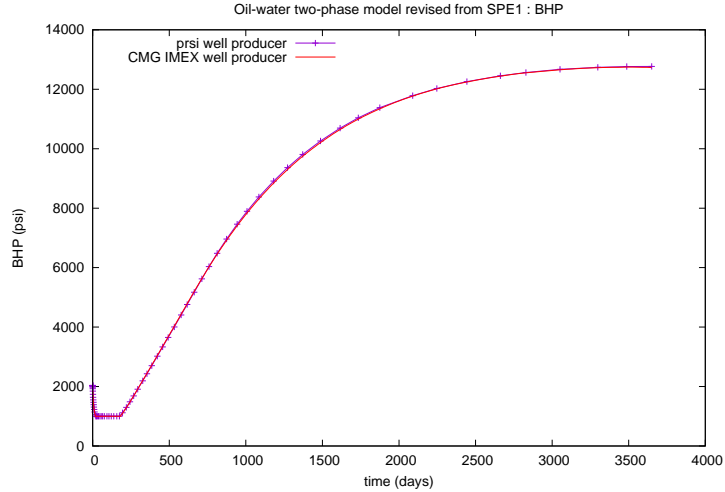

Figure 5: BHP for well injector and producer of example two-phase model from spe1 (pressure unit: psi).

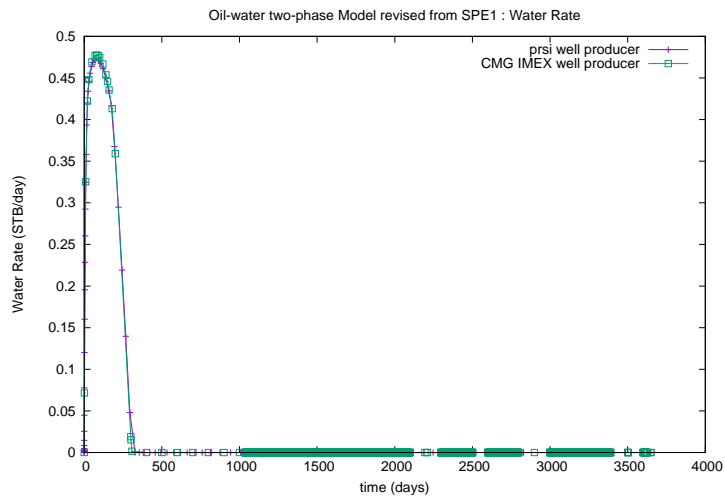

Figure 6: Water rate for well injector and producer of example two-phase model from spe1 (pressure unit: stb/day).

The results from this example are similar. We compare oil production rate, bottom hole pressure and water rate with CMG IMEX, which are shown in Figure 4, 5, 6, respectively. Again, we can see that our results match CMG IMEX, which prove our methods and implementation are correct.

### Newton Methods

This section compares the standard Newton method and the inexact Newton method.

**Example 3** *The SPE10 project[33] is applied. The stopping criterion for both nonlinear methods is 1e-2. The stopping criterion for the linear systems from the standard Newton method is 1e-4. The criterion for the linear systems from inexact Newton method is determined automatically by the simulator. Other settings are the same for these two methods. The maximal iterations for the nonlinear methods are 20.*

*The BICGSTAB method is the linear solver for all linear systems, and the maximal number of iterations for BICGSTAB is 300. The preconditioner is the CPR-FPF method. The overlap for RAS method is 1. The local potential reordering technique and the Quasi-IMPES strategy are applied. The simulation period is 2,000 days and the maximal time step is 100 days. The numerical summaries for the standard Newton method is shown in Table 2 and the summaries for the inexact Newton method is shown in Table 3. The numerical results are compared with other simulators to validate our implementation, including average pressure and oil rate, which are shown in Fig 7 and 8. Their scalability is shown by Figure 9.*

Table 2: Numerical summaries of the Newton method

| # procs | # Steps | # Ntn | # Slv | # Avg. solver | Time (s) | Avg. time (s) |
|---|---|---|---|---|---|---|
| 4  | 51 | 240 | 8817 | 36.7 | 102546.6 | 427.2 |
| 8  | 50 | 236 | 8916 | 37.7 | 46241.1  | 195.9 |
| 16 | 53 | 247 | 9258 | 37.4 | 23753.3  | 96.1 |
| 32 | 51 | 240 | 8789 | 36.6 | 11109.0  | 46.2 |
| 64 | 52 | 244 | 9393 | 38.4 | 6236.8   | 25.5 |

Table 3: Numerical summaries of the inexact Newton method

| # procs | # Steps | # Ntn | # Slv | # Avg. solver | Time (s) | Avg. time (s) |
|---|---|---|---|---|---|---|
| 4  | 88 | 522 | 5231 | 10.0 | 87708.3 | 168.0 |
| 8  | 91 | 542 | 5712 | 10.5 | 44873.2 | 82.7 |
| 16 | 84 | 499 | 5010 | 10.0 | 19840.4 | 39.7 |
| 32 | 96 | 582 | 6177 | 10.6 | 12024.4 | 20.6 |
| 64 | 92 | 551 | 5851 | 10.6 | 5585.6  | 10.1 |

Fig 7 compares average pressure of all grid cells and the average pressure is compared with a commercial simulator, from which we can see our results (`Avg Pres, UC`) match that simulator. Fig 8 compares oil rate with results from several oil companies. We can also see that our results (`UCalgary`) match other simulators very well.

In Tables 2 and 3, the number of MPI processors, number of time steps, total nonlinear iterations, total linear iterations, average linear iterations per Newton iteration (# Avg. solver), total simulation time and average time per Newton iteration (Avg. time) are presented. From these two tables, we can see that the standard Newton method requires much less time steps than the inexact Newton method, which means that the average time step of the standard Newton method is larger. The total number of the nonlinear iterations for the inexact Newton method is much larger than the standard Newton method. The results show that the standard Newton method converges faster than the inexact Newton method for this project. One reason is that the standard Newton method solves the linear systems much more accurately than the inexact Newton method. However, for the linear iterations, we can see that the standard Newton method requires three times more iterations than the inexact Newton method due to the accuracy required by the standard Newton method. As a consequence, the standard Newton method uses more simulation time than the inexact Newton method. Fig 9 shows that the parallel techniques applied in the manuscript are scalable and effective.

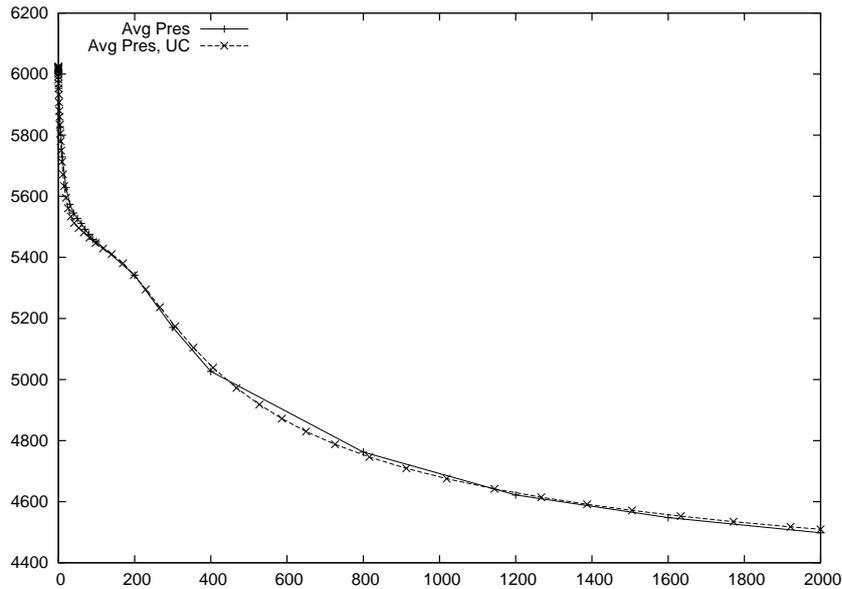

Figure 7: Comparison of average pressure.

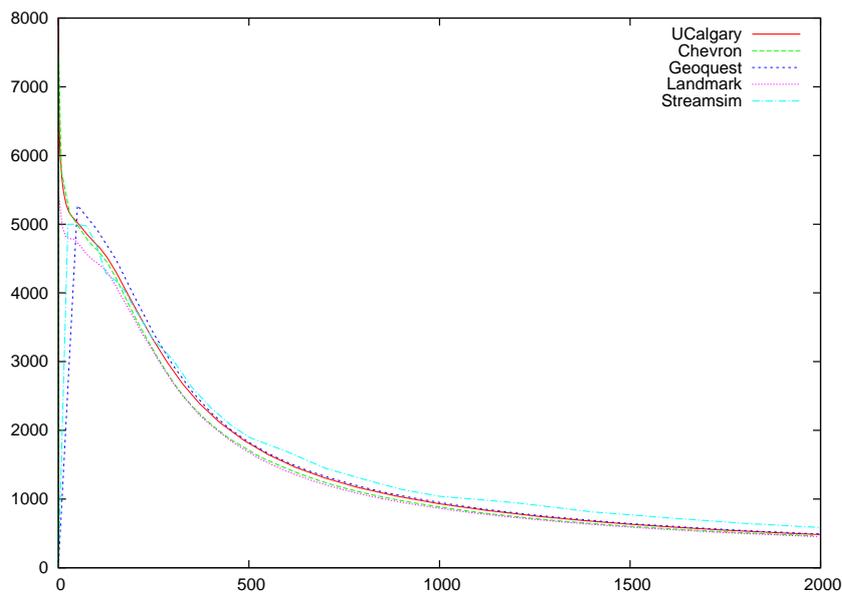

Figure 8: Comparison of oil rate.

## Matrix Processing

**Example 4** *The case is SPE10 project. The inexact Newton method is applied, its stopping criterion is 1e-2 and its maximal Newton iterations are 20. The restarted GMRES method is the linear solver for the solution of Jacobian systems. The overlap for the RAS method is 1. The maximal number of iterations for the solver is 100. The simulation period is 2,000 days and the maximal time step is 100 days. The numerical summaries are given in Table 4. The performance of different stages of the solver is presented in Table 5 and 64 MPI processors (CPU cores) are employed.*

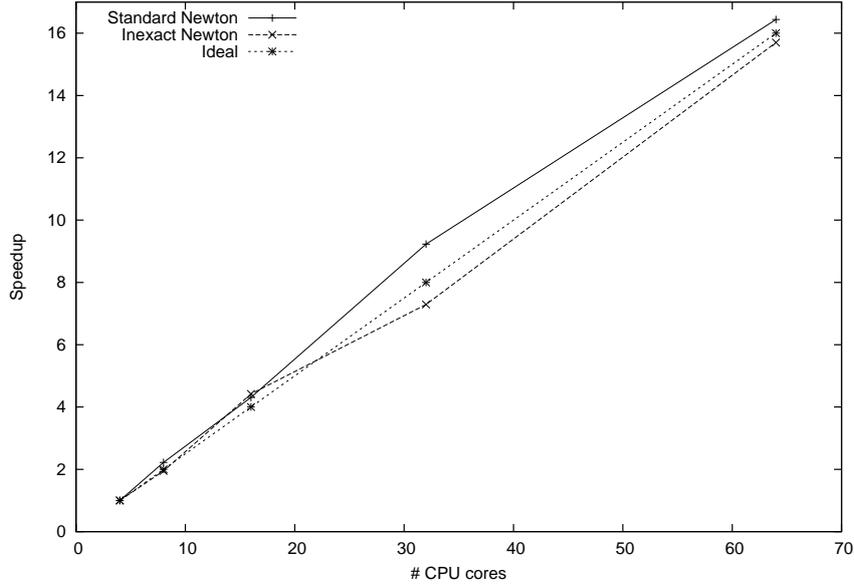

Figure 9: Scalability of the Newton and inexact Newton methods

Table 4: Numerical summaries of the matrix processing strategies

| # procs | # Steps | # Ntn | # Slv | # Avg. solver | Time (s) | Avg. time (s) |
|---------|---------|-------|-------|---------------|----------|---------------|
| NONE    | 161 (802.65) | 1808 | 401342 | 221.98 | 97784.1 | 54.1 |
| PT      | 198     | 1370  | 267182 | 195.0 | 74686.8 | 54.5 |
| ABF     | 166     | 1075  | 36043  | 33.5  | 14038.6 | 13.0 |
| QI      | 77      | 466   | 15965  | 34.2  | 5597.6  | 12.0 |
| PT-ABF  | 92      | 550   | 19930  | 36.2  | 8804.2  | 16.0 |
| PT-QI   | 71      | 403   | 12976  | 32.1  | 5437.0  | 13.4 |

When the potential reordering and the matrix decoupling methods are both disabled (NONE), the simulation fails, which indicates that the potential reordering and the matrix decoupling are necessary for the preconditioner. When only the potential reordering method is applied (PT), the simulation succeeds. However, the number of its linear iterations are 267,182 and the number of its average linear iterations per Newton iteration are 195. The result means that applying the potential reordering method alone is not effective enough. When using the Quasi-IMPES strategy only (QI), the numbers of the linear iterations and the nonlinear iterations are 15,965 and 466, respectively, which is much more efficient than the potential reordering method. When using the ABF strategy only (ABF), the number of the linear iterations are 36,043. When using both the potential reordering method and the Quasi-IMPES strategy (PT-QI), the numbers of the linear iterations and the nonlinear iterations are 12,976 and 403, respectively, which is better than using either the potential reordering method or the Quasi-IMPES strategy. When using both the potential reordering method and the ABF strategy (PT-ABF), its performance is also better than applying the ABF strategy. We can conclude that the potential reordering method and the matrix decoupling strategies are important to the solution of the Jacobian systems. For the Quasi-IMPES and ABF decoupling strategies, they have better performance when working with the potential reordering method.

Table 5: Performance of different stages of the linear solver and preconditioner

| # Nprocs | PT (s) | ABF (s) | QI (s) | RAS (s) | AMG (s) | Itr (s) |
|---|---|---|---|---|---|---|
| 32 | 0.696 | 1.254 | 0.258 | 1.048 | 2.908 | 0.635 |
| 64 | 0.364 | 0.628 | 0.138 | 0.622 | 1.309 | 0.309 |
| 128 | 0.193 | 0.306 | 0.063 | 0.354 | 0.665 | 0.159 |

Table 5 shows the performance of the locally potential reordering (PT), the ABF decoupling strategy (ABF), the Quasi-IMPES decoupling strategy (QI), the assembling of the RAS method (RAS), the assembling of the algebraic multigrid method (AMG) and one solver iteration (Itr). The results show that the locally potential reordering takes around two solver iterations and the running time of the locally potential reordering reduces if more MPI processors are applied. The ABF strategy requires much more time than the Quasi-IMPES strategy and they require less time if more MPI processors are employed. The assembling of the RAS method and the AMG method take around six solver iterations. The total cost of the assembling stage is around nine solver iterations, while Table 4 shows that the solver of each linear system converges within around 35 linear iterations. We can see that the assembling stage takes around 20% of the total solution time (including the assembling and solving of linear systems). The assembling stage and the solving stage are also scalable.

**Linear Solvers**

This section studies the linear solvers, where the restarted GMRES(m) method and the BICGSTAB method are tested.

**Example 5** *The case is SPE10 project. The inexact Newton method is applied and its stopping criterion is 1e-2. Its maximal Newton iterations are 20. The restarted GMRES(m) method is the GMRES(50) method. The maximal number of (inner) iterations for the solvers is 100. The overlap for the RAS method is 1. The potential reordering and the Quasi-IMPES strategies are applied. The simulation period is 2,000 days and the maximal time step is 100 days. The numerical summaries for GMRES(50) and BICGSTAB are presented in Tables 6 and 3, respectively.*

Table 6: Numerical summaries of the restarted GMRES(50) method

| # procs | # Steps | # Ntn | # Slv | # Avg. solver | Time (s) | Avg. time (s) |
|---|---|---|---|---|---|---|
| 4 | 68 | 380 | 11245 | 29.5 | 82730.2 | 217.7 |
| 8 | 69 | 389 | 11758 | 30.2 | 41324.1 | 106.2 |
| 16 | 68 | 383 | 11428 | 29.8 | 20109.3 | 52.5 |
| 32 | 69 | 393 | 11415 | 29.0 | 9875.1 | 25.1 |
| 64 | 71 | 403 | 12976 | 32.1 | 5438.2 | 13.4 |

Tables 6 and 3 show that the GMRES(m) method uses smaller time steps and fewer Newton iterations than the BICGSTAB method. However, the GMRES(m) method uses much more linear iterations than the BICGSTAB method. The numerical results show that both the GMRES(m) method and the BICGSTAB method are efficient in the simulations, whose average iterations per Newton iteration are around 30 and 10 for GMRES(50) and BICGSTAB, respectively. We should mention that the GMRES(m) solves one preconditioning system in each inner linear iteration, while the BICGSTAB method solves two preconditioning systems in each linear iteration. However, the overall running time for both methods is comparable. The average running time per Newton iteration also shows that they are scalable.

## Scalability

This section tests the strong scalability of the parallel techniques developed.

**Example 6** *This example benchmarks a refined SPE10 case, where each grid block is refined into eight grid blocks. This case has around nine millions of grid blocks and around 18 millions of unknowns. The stopping criterion for the inexact Newton method is 1e-3 and the maximal Newton iterations are 20. The BICGSTAB solver is applied and its maximal iterations are 100. The potential reordering and the Quasi-IMPES decoupling strategy are applied. The overlap for RAS method is 1. The simulation period is 20 days and the maximal time step is 10 days. The numerical summaries are shown in Table 7 and the speedup (scalability) is shown in Figure 10.*

Table 7: Numerical summaries of Example 6

| # procs | # Steps | # Ntn | # Slv | # Avg. solver | Time (s) | Avg. time (s) |
|---------|---------|-------|-------|---------------|----------|---------------|
| 16      | 37      | 231   | 2816  | 12.1          | 93900.4  | 406.4         |
| 32      | 35      | 205   | 2569  | 12.5          | 37854.6  | 184.6         |
| 64      | 38      | 225   | 3265  | 14.5          | 21699.5  | 96.4          |
| 128     | 38      | 244   | 3290  | 13.4          | 10514.2  | 43.0          |
| 256     | 37      | 230   | 3044  | 13.2          | 4889.9   | 21.2          |
| 512     | 40      | 248   | 3640  | 14.6          | 3071.8   | 12.3          |
| 1024    | 37      | 231   | 3177  | 13.7          | 1555.7   | 6.7           |

This case is more difficult than the original SPE10 problem. The results from Table 7 show that the simulations take around 40 time steps to complete 20 days, which means that the average time step is small. The average linear iterations for each Newton iteration are between 12 and 15, which are slightly more than that for the original SPE10 problem. The simulation times and Figure 10 show that the parallel techniques here have good scalability.

**Example 7** *This example tests another refined SPE10 case, where each grid block is refined into 27 grid blocks. This case has around 30 millions of grid blocks and around 60 millions of unknowns. The stopping criterion for the inexact Newton method is 1e-2 and the maximal Newton iterations are 20. The BICGSTAB solver is applied and its maximal iterations are 100. The potential reordering and the Quasi-IMPES decoupling strategy are applied. The simulation period is 10 days and the maximal time step is 5 days. The numerical summaries are shown in Tables 8 and 9, and the speedup (scalability) is shown in Figure 11.*

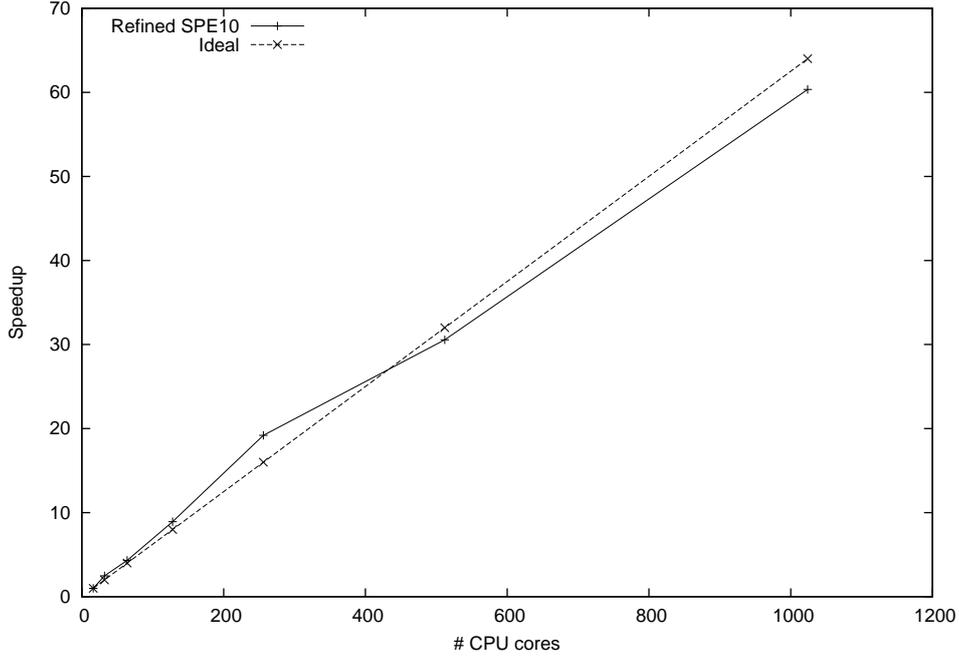

Figure 10: Scalability of Example 6

Table 8: Numerical summaries of Example 7, overlap 1

| # procs | # Steps | # Ntn | # Slv | # Avg. solver | Time (s) | Avg. time (s) |
|---------|---------|-------|-------|---------------|----------|---------------|
| 64      | 47      | 262   | 2569  | 9.8           | 94344.7  | 360.0         |
| 128     | 49      | 290   | 2710  | 9.3           | 45517.1  | 156.9         |
| 256     | 43      | 247   | 2506  | 10.1          | 20313.8  | 82.2          |
| 512     | 47      | 286   | 2606  | 9.1           | 11222.0  | 39.2          |
| 1024    | 47      | 278   | 2965  | 10.6          | 6177.3   | 22.2          |
| 2048    | 47      | 274   | 2845  | 10.3          | 3217.7   | 11.7          |

Table 9: Numerical summaries of Example 7, overlap 2

| # procs | # Steps | # Ntn | # Slv | # Avg. solver | Time (s) | Avg. time (s) |
|---------|---------|-------|-------|---------------|----------|---------------|
| 64      | 47      | 263   | 2153  | 8.1           | 89385.5  | 339.8         |
| 128     | 45      | 276   | 2312  | 8.3           | 42387.0  | 153.5         |
| 256     | 46      | 272   | 2506  | 9.2           | 22038.1  | 81.0          |
| 512     | 46      | 260   | 2412  | 9.2           | 10963.7  | 42.1          |
| 1024    | 45      | 253   | 2457  | 9.7           | 5631.0   | 22.2          |
| 2048    | 46      | 257   | 2486  | 9.6           | 3138.1   | 12.2          |

For this example, up to 2,048 MPI processors are employed. Tables 8 and 9 show numerical summaries of the simulations with overlaps one and two, respectively. The total number of time steps is between 40 and 50 and the average time step is smaller than that in Example 6 due to the smaller size of grid blocks.

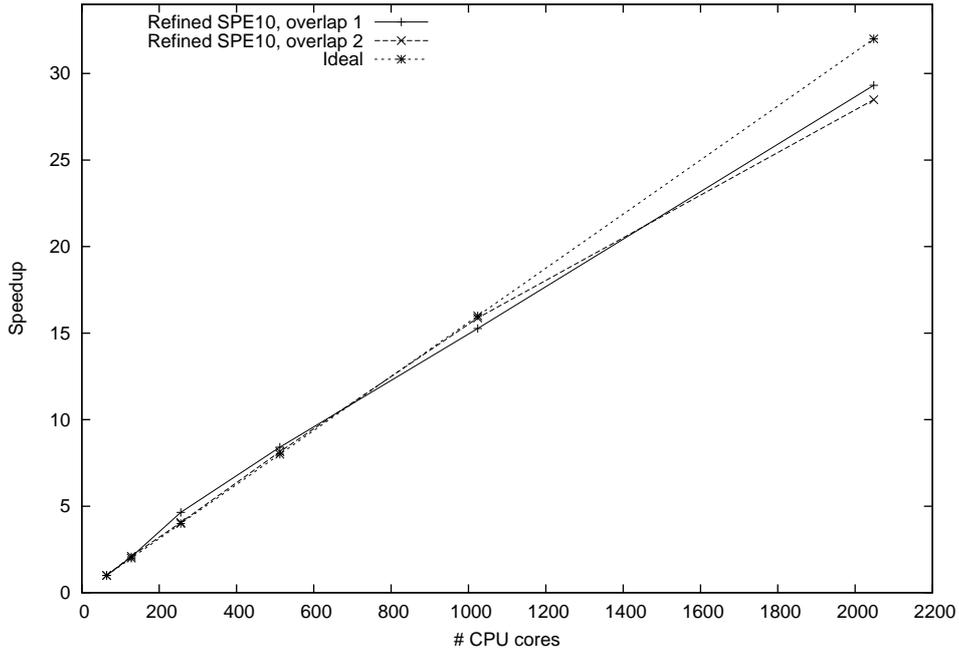

Figure 11: Scalability of Example 7

The tables show that the linear solver converges in around 10 iterations, which means that both the linear solver and the preconditioner are efficient. When we increase the number of MPI processors, the average linear iterations per Newton iteration increase slightly, which demonstrates the robustness of the linear solver and the preconditioner. The running times and Figure 11 show good scalability; especially when the number of MPI processors is not larger than 1,024, the scalability is almost ideal. For the case with 2,048 MPI processors, each node (computer card) runs 16 MPI processors, which share the memory and network, and compete the computation and communication resources, the scalability is reduced slightly, and the running time is slightly longer than that in the ideal condition. We emphasize that the results show our parallel techniques and methods are efficient, robust and scalable.

## CONCLUSIONS

This paper studies two-phase oil-water model and various numerical techniques are studied, including nonlinear methods, linear solvers and advanced matrix processing strategies, which are mainly developed for distributed memory parallel systems. Therefore, their effectiveness and scalability are important. The matrix decoupling strategies are applied locally and no communication is involved. In this case, the scalability is ideal. For the matrix reordering, the potential reordering strategy is applied, which is applied locally and so is the matrix decoupling strategy. The communication volume for the matrix reordering technique is small and this technique is also scalable for parallel systems. Numerical experiments for the nonlinear methods, matrix processing techniques, linear solvers, and RAS method show that these methods and techniques are efficient and scalable. The oil-water model is a simpler model for reservoir simulation compared with the black oil, polymer flooding, compositional and thermal models. However, those numerical techniques can be applied to these more complex models directly.

# ACKNOWLEDGEMENTS

The support of Department of Chemical and Petroleum Engineering, University of Calgary and Reservoir Simulation Research Group is gratefully acknowledged. The research is partly supported by NSERC, AIEES, Foundation CMG, AITF iCore, IBM Thomas J. Watson Research Center, and the Frank and Sarah Meyer FCMG Collaboration Centre for Visualization and Simulation. The research is also enabled in part by support provided by WestGrid (www.westgrid.ca), SciNet (www.scinethpc.ca) and Compute Canada Calcul Canada (www.computecanada.ca).

# References

[1] K. Li, A. Dogru, A. McDonald, A. Merchant, A. Al-Mulhem, S. Al-Ruwaili, N. Sobh, and H. Al-Sunaidi, Improving the Performance of Mars Simulator on Cray-2 Supercomputer, Middle East Oil Show, Society of Petroleum Engineers, Bahrain, 11-14 March 1995.

[2] J. Wheeler, and R. Smith, Reservoir Simulation on a Hypercube, SPE Reservoir Engineering, 1990, 5(04), 544.

[3] J. Rutledge, D. Jones, W. Chen, and E. Chung, The Use of Massively Parallel SIMD Computer for Reservoir Simulation, SPE Computer Applications, 1992, 4(04), 16.

[4] G. Shiralkar, R. Stephenson, W. Joubert, O. Lubeck, and B. van Bloemen Waanders, Falcon: A production quality distributed memory reservoir simulator, SPE Reservoir Simulation Symposium, Society of Petroleum Engineers, Dallas, 8-11 Jun 1997.

[5] M. Parashar, J. Wheeler, G. Pope, K. Wang, and P. Wang, A New Generation EOS Compositional Reservoir Simulator: Part II, Framework and Multiprocessing, SPE Reservoir Simulation Symposium, Society of Petroleum Engineers, Dallas, 8-11 Jun 1997.

[6] T. Kaarstad, J. Froyen, P. Bjorstad, M. Espedal, A Massively Parallel Reservoir Simulator, SPE Reservoir Simulation Symposium, Society of Petroleum Engineers, San Antonio, 12-15 Feb 1995.

[7] J. Killough, D. Camilleri, B. Darlow, J. Foster, Parallel Reservoir Simulator Based on Local Grid Refinement, SPE Reservoir Simulation Symposium, Society of Petroleum Engineers, Dallas, 8-11 Jun 1997.

[8] A. Dogru, H. Sunaidi, L. Fung, W. Habiballah, N. Al-Zamel, K. Li, A parallel reservoir simulator for large-scale reservoir simulation. SPE Reservoir Evaluation & Engineering, 2002, 5(1), 11.

[9] A. Dogru, L. Fung, U. Middya, T. Al-Shaalan, J. Pita, A next-generation parallel reservoir simulator for giant reservoirs, SPE/EAGE Reservoir Characterization & Simulation Conference, Society of Petroleum Engineers, The Woodlands, 2-4 Feb 2009.

[10] Y. Saad, Iterative methods for sparse linear systems, 2nd edition, SIAM, Philadelphia, 2003.

[11] J. Wallis, Incomplete Gaussian elimination as a preconditioning for generalized conjugate gradient acceleration, SPE Reservoir Simulation Symposium, Society of Petroleum Engineers, San Francisco, 15-18 Nov 1983.

[12] G. Behie, and P. Forsyth, Incomplete factorization methods for fully implicit simulation of enhanced oil recovery. SIAM Journal on Scientific and Statistical Computing, 1984, 5(3), 543.

[13] J. Wallis, R. Kendall and T. Little, Constrained residual acceleration of conjugate residual methods, SPE Reservoir Simulation Symposium, Society of Petroleum Engineers, Dallas, 10-13 Feb 1985.

[14] H. Cao, H. Tchelepi, J. Wallis, H. Yardumian, Parallel scalable unstructured CPR-type linear solver for reservoir simulation. SPE Annual Technical Conference and Exhibition, Society of Petroleum Engineers, Dallas, 9-12 Oct 2005.

[15] R. Bank, T. Chan, W. Coughran Jr., R. Smith, The Alternate-Block-Factorization procedure for systems of partial differential equations, BIT Numerical Mathematics, 1989, 29(4), 938.

[16] K. Stüben, T. Clees, H. Klie, B. Lou, M. Wheeler, Algebraic multigrid methods (AMG) for the efficient solution of fully implicit formulations in reservoir simulation, SPE Reservoir Simulation Symposium, Society of Petroleum Engineers, Houston, 26-28 Feb 2007.

[17] K. Stüben, A review of algebraic multigrid, Journal of Computational and Applied Mathematics, 2001, 128(1), 281.

[18] S. Lacroix, Y. Vassilevski, and M. Wheeler, Decoupling preconditioners in the implicit parallel accurate reservoir simulator (IPARS), Numerical linear algebra with applications, 2001, 8(8), 537.

[19] R. Scheichl, R. Masson, and R. Wendebourg, Decoupling and Block Preconditioning for Sedimentary Basin Simulations, Computational Geosciences, 2003, 7, 295.

[20] T. Al-Shaalan, H. Klie, A. Dogru, M. Wheeler, Studies of Robust Two Stage Preconditioners for the Solution of Fully Implicit Multiphase Flow Problems, SPE Reservoir Simulation Symposium, Society of Petroleum Engineers, The Woodlands, 2-4 Feb 2009.

[21] X. Hu, W. Liu, G. Qin, J. Xu, C. Zhang, Development of a fast auxiliary subspace pre-conditioner for numerical reservoir simulators, SPE Reservoir Characterisation and Simulation Conference and Exhibition, Society of Petroleum Engineers, Abu Dhabi, 9-11 Oct 2011.

[22] H. Liu, K. Wang, Z. Chen, and K. Jordan, Efficient Multi-stage Preconditioners for Highly Heterogeneous Reservoir Simulations on Parallel Distributed Systems, SPE Reservoir Simulation Symposium, Society of Petroleum Engineers, Houston, 23-25 Feb 2015.

[23] T. Chen, N. Gewecke, Z. Li, A. Rubiano, R. Shuttleworth, Y. Yang, and X. Zhong, Fast Computational Methods for Reservoir Flow Models, report-082009, University of Minnesota, 2009.

[24] A. Toselli, and O. Widlund, Domain decomposition methods: algorithms and theory, Vol 34, Springer, New York, 2005.

[25] L. Zhang, T. Cui, and H. Liu, A set of symmetric quadrature rules on triangles and tetrahedra, J. Comput. Math, 2009, 27(1), 89.

[26] X. Cai, and M. Sarkis, A restricted additive Schwarz preconditioner for general sparse linear systems, SIAM Journal on Scientific Computing, 1999, 21(2), 792.

[27] Z. Chen, G. Huan, and Y. Ma, Computational methods for multiphase flows in porous media, Vol 2, SIAM, Philadelphia, 2006.

[28] F. Kwok F, and H. Tchelepi, Potential-based reduced newton algorithm for nonlinear multiphase flow in porous media, Journal of Computational Physics, 2007, 227(1), 706.

[29] R. Falgout, and U. Yang, Hypre: A library of high performance preconditioners, Lecture Notes in Computer Science, 2002, 632.

[30] H. Liu, K. Wang, Z. Chen, K. Jordan, J. Luo, and H. Deng, A Parallel Framewrok for Reservoir Simulators on Distributed-memory Supercomputers, SPE/IATMI Asia Pacific Oil & Gas Conference and Exhibition, Society of Petroleum Engineers, Nusa Dua, 20-22 Oct 2015.

[31] H. Liu, Dynamic Load Balancing on Adaptive Unstructured Meshes, 10th IEEE International Conference on High Performance Computing and Communications, Dalian, 25-27 Sept 2008.

[32] A. Sedighi, Y. Deng, B. Zhang, Fairness of Task Scheduling in High Performance Computing Environments, Scalable Computing: Practice and Experience, 2014, 15(3), 273.

[33] M. Christie, and M. Blun, Tenth SPE comparative solution project: A comparison of upscaling techniques. SPE Reservoir Evaluation & Engineering, 2001, 4(4), 308.

[34] G. Karypis, K. Schloegel, and V. Kumar, Parallel static and dynamic multi-constraint graph partitioning, Concurrency and Computation: Practice and Experience, 2002, 14(3), 219.

[35] L. Zhang, A Parallel Algorithm for Adaptive Local Refinement of Tetrahedral Meshes Using Bisection, Numer. Math.: Theory, Methods and Applications, 2009, 2, 65.